\documentclass[a4paper, aps, prb, amsmath, amssymb, superscriptaddress, preprintnumbers, onecolumn, 12pt]{revtex4-1}
\usepackage{graphicx,xcolor}
\usepackage{epstopdf}
\usepackage{mathptmx, textcomp}
\usepackage[utf8]{inputenc}
\usepackage[T1]{fontenc}
\usepackage{braket}
\usepackage{amsmath}
\usepackage{upgreek}
\usepackage{color}
\usepackage{xcolor} 
\usepackage[UKenglish]{babel}
\usepackage{array}

\def \bk{{\bf k}}

\def \mB{\mathrm{B}}

\def \mm{\mathrm{m}}

\def \mtot{\mathrm{tot} }

\def \mHz{\mathrm{Hz}}
\def \mnK{\mathrm{nK}}

\begin{document}

\title{Second sound in the BEC-BCS crossover}
	
\author{Daniel K. Hoffmann}
\affiliation{Institut f\"{u}r Quantenmaterie and Center for Integrated Quantum Science
and Technology (IQST), Universit\"{a}t Ulm, D-89069 Ulm, Germany}
\author{Vijay Pal Singh}
\affiliation{Institut f\"{u}r Laserphysik, Zentrum f\"{u}r Optische Quantentechnologien and The Hamburg center for Ultrafast Imaging, Universit\"{a}t Hamburg, Luruper Chaussee 149, 22761 Hamburg, Germany}
\author{Thomas Paintner}
\affiliation{Institut f\"{u}r Quantenmaterie and Center for Integrated Quantum Science
and Technology (IQST), Universit\"{a}t Ulm, D-89069 Ulm, Germany}
\author{Manuel Jäger}
\affiliation{Institut f\"{u}r Quantenmaterie and Center for Integrated Quantum Science
and Technology (IQST), Universit\"{a}t Ulm, D-89069 Ulm, Germany}
\author{Wolfgang Limmer}
\affiliation{Institut f\"{u}r Quantenmaterie and Center for Integrated Quantum Science
and Technology (IQST), Universit\"{a}t Ulm, D-89069 Ulm, Germany}
\author{Ludwig Mathey}
\affiliation{Institut f\"{u}r Laserphysik, Zentrum f\"{u}r Optische Quantentechnologien and The Hamburg center for Ultrafast Imaging, Universit\"{a}t Hamburg, Luruper Chaussee 149, 22761 Hamburg, Germany}
\author{Johannes Hecker Denschlag*}
\affiliation{Institut f\"{u}r Quantenmaterie and Center for Integrated Quantum Science
and Technology (IQST), Universit\"{a}t Ulm, D-89069 Ulm, Germany}

\vspace{0.3cm}
\begin{abstract}
\section*{Abstract}
Second sound is an entropy wave which propagates in the superfluid component of a quantum liquid. 
Because it is an entropy wave, it probes the thermodynamic properties of the quantum liquid which are determined, e.g., by the interaction strength between the particles of the quantum liquid and their temperature. 
Here, we study second sound propagation for a large range of interaction strengths within the crossover between a Bose-Einstein condensate (BEC) and the Bardeen-Cooper-Schrieffer (BCS) superfluid. 
In particular, we investigate the strongly-interacting regime where currently theoretical predictions only exist in terms of an interpolation between the BEC, BCS and unitary regimes. 
Working with a quantum gas of ultracold fermionic $^6$Li atoms with tunable interactions, we show that the second sound speed varies only slightly in the crossover regime.
We gain deeper insights into sound propagation and excitation of second s ound by varying the excitation procedure which ranges from a sudden force pulse to a gentle heating pulse at the cloud center. 
These measurements are accompanied by classical-field simulations which help with the interpretation of the experimental data.
Furthermore, we determine the spatial extension of the superfluid phase and estimate the superfluid density.  
In the future, this may be used to construct the so far unknown equation of state throughout the crossover. 
\end{abstract}

\maketitle
\section*{Introduction}
Second sound is a transport phenomenon of quantum liquids that emerges below the critical temperature for superfluidity $T_C$ \cite{Hal18, Don09,Pit15}. 
It was experimentally discovered\cite{Pes46} in 1944 in He II\cite{Gri09} and was described with a hydrodynamic two-fluid model\cite{Don09,Tis38,Lan41,Put74} which treats He II as a mixture of a superfluid (SF) and a normal fluid (NF).
The SF component has no entropy and  flows without dissipation. 
The NF component carries all the entropy and has non-zero viscosity. 
In the limit of vanishing temperature $T \rightarrow 0$, the two-fluid model predicts that
first sound (i.e. standard sound waves) correspond to a propagating pressure oscillation with constant entropy, while second sound is an entropy oscillation propagating at constant pressure \cite{Put74}.

The properties of a superfluid naturally depend on parameters such as its temperature and the interaction strength between its particles. 
With the advent of ultracold quantum gases, with tunable interactions, these dependencies can now be studied. 
In particular, an ultracold fermionic quantum gas with a tunable Feshbach resonance offers a unique opportunity to access various sorts of superfluidity in one system, ranging continuously between a Bose-Einstein condensate (BEC) of bosonic molecules, a resonant superfluid, and a superfluid gas of Cooper pairs (BCS superfluid) \cite{Gio08,Ket08,Zwe12}.
In the experiment this is done by tuning the interaction parameter $(k_{\rm{F}}a)^{-1}$, where $a$ is the scattering length, $k_F=\sqrt{2mE_F}/\hbar$ the Fermi wavenumber, $E_F$  is the Fermi energy and $m$ the atomic mass.

A large range of thermodynamical properties of the BEC-BCS crossover has been studied e.g. in refs.\cite{Nas10,Ku12,Sh08,Jin15,Jos07,Tay09,Per04,Zwe12}.
Recently, second sound has been measured by Sidorenkov et al.\cite{Sid13} in a unitary Fermi gas and  by Ville et al. \cite{Vil18} in a two-dimensional bosonic superfluid.

Here, we experimentally investigate how second sound changes across the BEC – BCS crossover. This is especially important, since full theoretical calculations are not yet available in the strongly interacting regime. Nevertheless, comparing our measurements to existing calculations and interpolations we find reasonable agreement.  
In particular, c-field simulations in the BEC regime match quite well the corresponding observed wave dynamics of the experiment up to an interaction strength of $1/k_Fa = 1$.

Furthermore, we explore how to tune second sound generation by testing experimentally and theoretically various excitation schemes ranging from a gentle local heating of the superfluid to a short local force pulse. 
As second sound is mainly an entropy wave and first sound is mainly a pressure wave, these different excitation schemes give rise to different responses for first and second sound. 
This helps for separating the generally weak second sound signals from the first sound ones. 
We find, that this separation works especially well when both first and second sound are excited as density dip wavepackets. 
For this case, we were able to quantitatively compare the amplitudes of first and second sound and compare the results to a prediction. 

\section*{Results}

\subsection*{Experimental details}
Our experiments are carried out with a balanced, two-component ultracold gas of fermionic $^6$Li atoms in the two lowest hyperfine states $\ket{F,m_F}=\ket{1/2,\pm1/2}$ of the electronic ground state. 
The gas is confined by a combined magnetic and optical dipole trap with a trap depth of $U_0 \approx 1\,\upmu\mathrm{K}\times k_B$, for details see Ref. \cite{Pai18,Hof18}.
The trap is nearly harmonic and cylindrically symmetric with trapping frequencies $\omega_r=2\pi\times305\,\mathrm{Hz}$ and $\omega_{x}=2\pi\times21\,\mathrm{Hz}$.
The temperature and the particle density are controlled by evaporative cooling. 
In the experiments the temperature ranges approximately from $0.12\,T_{\mathrm{F}}$ to $0.28\,T_{\mathrm{F}}$, where $T_\mathrm{F}=E_F/k_\mathrm{B}= \hbar(3 \bar \omega^3 N)^{1/3}$ is the Fermi temperature, $\bar{\omega}=\left(\omega_{x}\omega_r^2\right)^{1/3}$ is the geometric mean of the trapping frequencies and $N$ is the total number of atoms. 
The scattering length $a$ is tunable with an external magnetic field $B$ via a magnetic Feshbach resonance at $832\,\mathrm{G}$ \cite{Zur13}.

\begin{figure}[hbp]
\includegraphics[trim=0 0 0 0,width=1\textwidth]{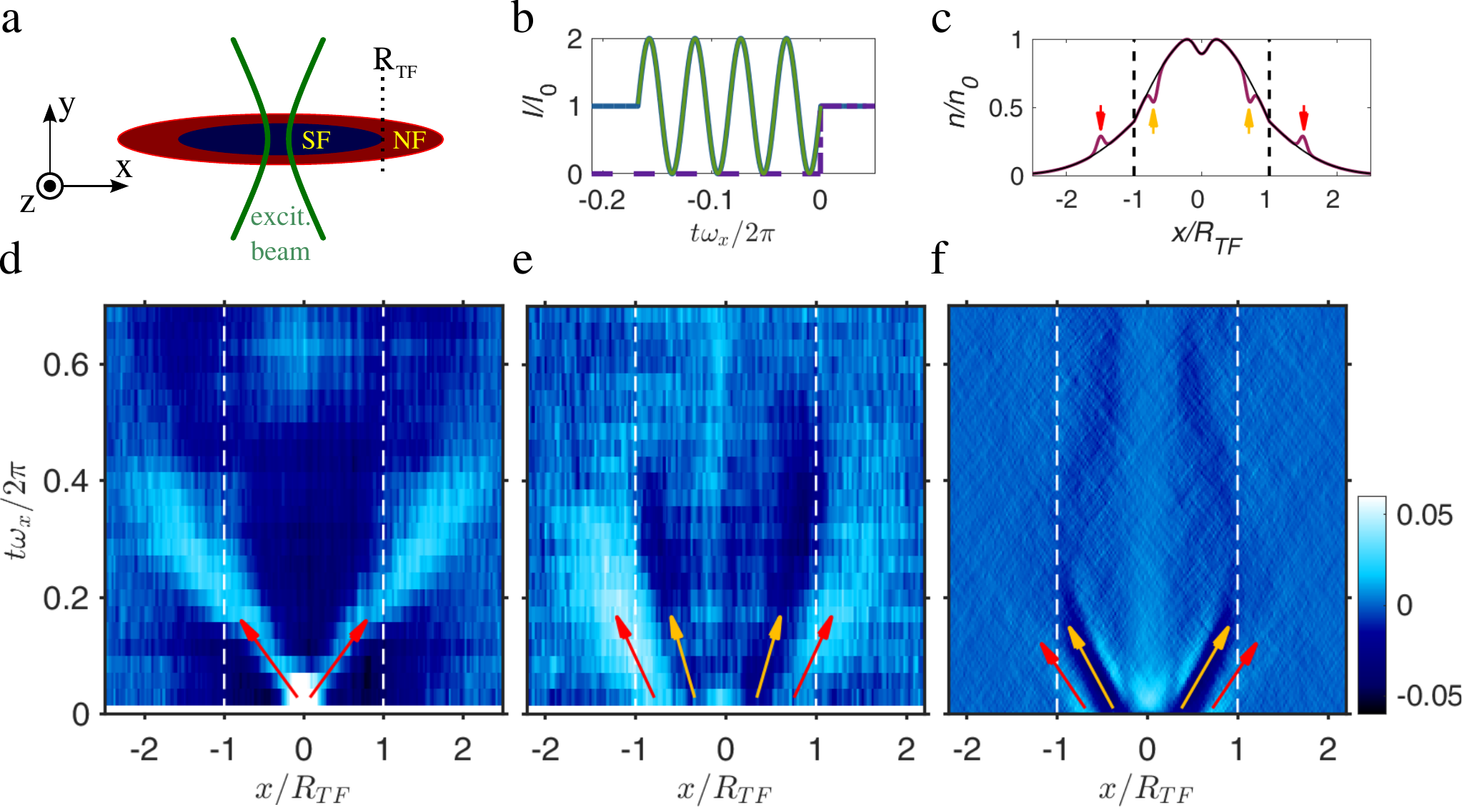}
\caption{\textbf{Sound excitation in a trapped superfluid Fermi gas in the vicinity of the BEC-BCS crossover.} \textbf{a,} Set-up:  A focussed, intensity-modulated, blue-detuned laser beam excites sound waves in the cigar-shaped atom cloud. \textbf{b,} Two different modulation sequences of the laser intensity. Purple dashed line: step excitation. Green solid line: heat pulse. The time $t$ is given in units of the axial trapping period $2\pi/\omega_{x}$. \textbf{c,} Sketch of a bimodal density distribution of a trapped BEC (purple line) at $y=z=0$. At the center of the trap a blue detuned beam produces a dimple in the potential. Modulating the beam intensity produces first sound waves (red arrows) and second sound (orange arrows) waves. Second sound reduces the local density of the cloud, while for first sound a density peak emerges.  The thin black line shows the profile of the unperturbed cloud. \textbf{d,} The false color plot shows the measured local change in the density $\Delta \bar{n}(x,t)$ as a function of axial position $x$ and time $t$. Here, $(k_Fa)^{-1}=(1.91\pm0.05)$ at  $B=735\,\mathrm{G}$ and $T/T_\mathrm{C}=(0.71\pm0.15)$. After excitation, two  wave packets (bright traces, marked with red arrows) propagate with first sound velocity $u_1$ towards the edges of the cloud. The excitation method predominantly excites first sound. Second sound is present as well but is barely discernible here. \textbf{e,} Propagation of first sound waves (bright traces, marked with red arrow) and second sound waves (dark traces, marked with orange arrows) after excitation with sinusoidal pulse of b). All other settings are the same as in d). \textbf{f,} Simulated sound propagation for the same parameters as in e). The orange arrows mark the propagating second sound and the red arrows the first sound, respectively.}
\label{fig:ComparisonTheoryExp}
\end{figure}

To excite sound modes in the system, we focus a blue-detuned $532\,\mathrm{nm}$ laser onto the trap center (see Ref.\cite{Sid13} and Fig.\,\ref{fig:ComparisonTheoryExp}a). 
The laser beam is aligned perpendicularly to the optical dipole trap and produces a repulsive potential barrier of $U_{ex}\approx0.2\,U_0$. 
At its focus, the beam has a waist of about $20\,\mathrm{\upmu m}$, which is comparable to the cloud size in the radial direction. 
To excite sound waves, the height of this additional potential is modulated. 
The excited sound modes generally exhibit contributions from both first and second sound \cite{Hei06,Ara09,Hu10}. 
However, it is possible to generate preferentially either one of the two sound modes by adapting the excitation method.

To excite primarily first sound, we abruptly switch on the excitation laser beam (see Fig. \ref{fig:ComparisonTheoryExp}b), similarly as for the first experiments on sound propagation in a dilute BEC \cite{And97}. 
This applies pressure on the cold cloud on both sides of the laser beam and creates two density wave packets (see Fig.\ref{fig:ComparisonTheoryExp}c) which propagate out in opposite directions along the axial trap axis with the speed $u_1$. 
In the experiments we detect these waves with the help of absorption imaging by measuring the density distribution of the atomic cloud as a function of time.

Figure\,\ref{fig:ComparisonTheoryExp}d shows such density waves for an experiment at $(k_Fa)^{-1}\approx (1.91\pm0.05)$, $B=735\,\mathrm{G}$ and a temperature of $T=(140\pm30)\,\mathrm{nK}=(0.28\pm0.06)\,T_F$, which corresponds to $T=(0.71\pm0.15)\,T_C$, where $T_\mathrm{C}$ is the critical temperature. 
For the given interaction strength, we used $T_C=0.4\,T_F$ (see Supplementary Note 1).  

Figure \ref{fig:ComparisonTheoryExp}d is a time ordered stack of  one-dimensional column density profiles of the atom cloud (see Methods for details). 
It shows the propagation of the sound waves along the axial direction $x$ as a function of time. 
The two density wave packets propagate with first sound velocity from the trap center towards the edge of the cloud (two bright traces, marked with red arrows). 
To obtain the speed of sound, we examine how the center position of each wave packet changes with time. 
The center positions are determined via a Gaussian fit. 
From Fig.\ref{fig:ComparisonTheoryExp}d we obtain $u_1=(17.2\pm3)\,\mathrm{mm/s}$ near the trap center. 
Our analysis shows that the sound propagation slows down as the pulse approaches the edge of the cloud where the particle density decreases. 
In the following, we focus on the sound speed close to the trap center.

To primarily excite second sound, we sinusoidally modulate the intensity of the excitation beam for $7\,\mathrm{ms}$ with a modulation frequency of $\omega_{ex}=2\pi \times 570\,\mathrm{Hz}\approx 2\omega_r$ and a modulation amplitude of $\Delta U\approx0.2\,U_0$. 
This parametrically heats the gas in radial direction (see Fig.\,\ref{fig:ComparisonTheoryExp}b).  
Subsequent thermalization via collisions occurs within a few milliseconds.
This creates a local depletion of the superfluid density, filled with normal gas, forming a region of increased entropy (see Fig.\ref{fig:ComparisonTheoryExp}c). 
This gives rise to two wave packets which propagate outwards along the axial direction with the speed of second sound.
Figure \ref{fig:ComparisonTheoryExp}e shows corresponding experimental data where we measure the local density distribution as in Fig. \ref{fig:ComparisonTheoryExp}d.
The second sound wave appears here as a density dip (dark traces, marked with orange arrows). 
A clear indication that the dark trace corresponds to second sound is the fact that it vanishes at the Thomas-Fermi radius $R_{TF}\approx 110\,\upmu{\rm m}$ where the superfluid fraction vanishes.
Second sound only propagates inside the superfluid phase.

Besides a second sound wave the excitation also produces a first sound wave (bright traces, marked with red arrows) which  propagates faster than the second sound wave and travels beyond the Thomas-Fermi radius.
The first sound wave is broader than in Fig.\,\ref{fig:ComparisonTheoryExp}d, which can be mainly explained by the longer excitation pulse.
To obtain $u_2$ we measure the time-dependent position of the minimum of each dark trace, which is deter\-mined via a Gaussian fit. 
For Fig.\,\ref{fig:ComparisonTheoryExp}e we obtain $u_2=(5.1\pm1.1)\,\mathrm{mm/s}$.

Figure\,\ref{fig:ComparisonTheoryExp}f shows numerical simulations of our experiment applying a dynamical c-field method\cite{Sin16} (see Supplementary Note 2 for detailed information on the method). 
The dimer scattering length\cite{Pet03} is $a_{dd}=0.6a$ and we assume all fermionic atoms to be paired up in molecules. 
To compare the simulations with the experimental results we choose the same values of $(k_{\mathrm{F}}a)^{-1}$ and the same central density of the trapped gas as in the experiment.  
The theory value for $u_2$ is $(5.7\pm0.05)\,\mathrm{mm/s}$ in agreement with the experimental value $(5.1\pm1.1)\,\mathrm{mm/s}$. 

\subsection*{Interaction strength dependence of second sound}

\begin{figure}[hbp]
\includegraphics[trim=0 0 0 0,width=.6\textwidth]{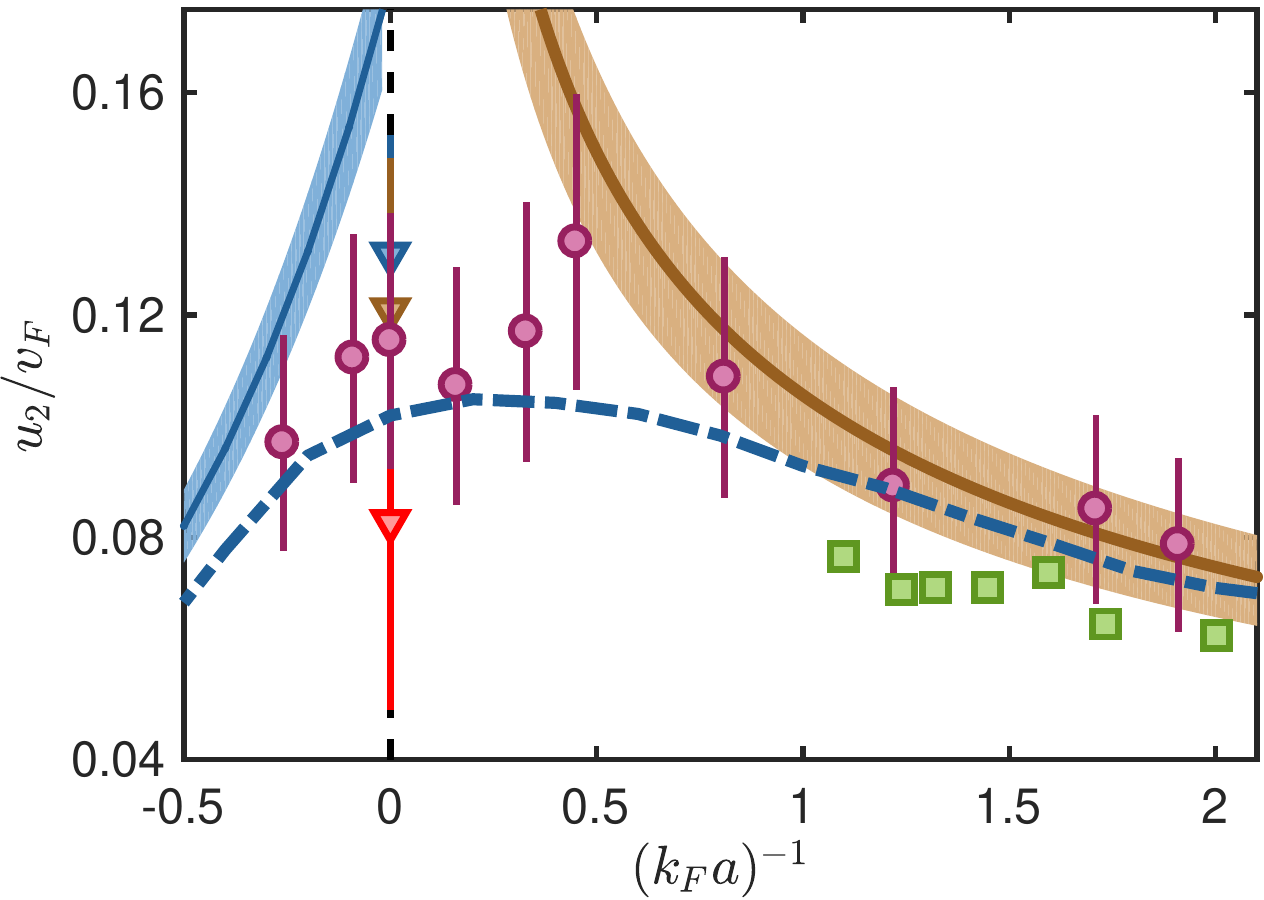}
\caption{\textbf{Second sound velocity $u_2$ as a function of interaction strength.}
The purple circles depict measured data for temperatures in the range
$T=65-145\,\mathrm{nK}$ which corresponds to $T/T_C=0.69-0.81$ (see Supplementary Note 1). The error bars are due to statistical uncertainties. The brown and blue solid line show hydrodynamic predictions for the BEC and BCS regime at $T=0.75\,T_C$, respectively (see Supp. Note 3). The shaded areas mark the second sound velocity in the temperature range of the experiments. The blue dash-dotted line shows a theoretical prediction of second sound in the crossover\cite{Hei06} for a homogeneous gas at $T/T_{\mathrm{C}}=0.75$. It interpolates between the results from hydrodynamic theory in the BEC and BCS regime. 
The green squares are results of our numerical c-field simulations which are consistent with both, analytic and experimental results. 
For comparison we also show the second sound velocity on the resonance measured in Ref. \cite{Sid13} at the temperatures $T/T_{\mathrm{C}}=0.65$ (blue triangle), $T/T_{\mathrm{C}}=0.75$ (brown triangle), and $T/T_{\mathrm{C}}=0.85$ (red triangle). }
\label{fig:FieldDep}
\end{figure}

We now perform measurements of second sound in the range $(-0.26\pm0.04)<(k_Fa)^{-1}<(1.91\pm0.05)$ of the BCS-BEC crossover. 
These are shown in Fig.\,\ref{fig:FieldDep} along with theoretical predictions. 
The second sound velocity $u_2$ is given in units of the Fermi velocity $v_F=\hbar k^{\mathrm{hom}}_F/m$. 
Here, the Fermi wavenumber $k^{\mathrm{hom}}_F$ is determined from the peak density at the trap center $k^{\mathrm{hom}}_F=\left(3\pi^2n_0\right)^{1/3}$. 
The blue dash-dotted line is a calculation from Ref.\cite{Hei06}, based on a hydrodynamic description in a homogeneous gas for the limiting cases of the BEC and the BCS regime, and unitarity. 
To connect these regimes, the results are interpolated across the crossover, bridging the range $\left|\left(k_Fa\right)^{-1}\right|<1$. 
The blue solid and the brown solid lines are our analytic hydrodynamic calculations which are valid in the BCS and BEC limit, respectively (see Supplementary Note 3).
For comparison, we show the results of the numerical c-field simulations (green squares), which agree with both, analytic description and experimental results.
Despite the large error bars the measurements indicate an increase of $u_2$ when approaching unitarity from the BEC side, in agreement with the theoretical results.

In general, second sound can only propagate in the superfluid phase of the gas. 
It is therefore natural to ask how the superfluid density $n_s$ and the speed of second sound $u_2$ are related. 
This relation could, in principle, be derived from the equation of state. 
However the equation of state is unknown for the strongly interacting regime. 
Nevertheless, we can still get a handle on the relationship between $n_s$ and $u_2$, by estimating the superfluid density for the regime of intermediate coupling, $1/k_F\,a > 1.5$, as follows. 
We carry out self-consistent mean-field calculations to determine the density distributions of the superfluid and the normal fluid for an interacting BEC in the trap (see Supplementary Note 4). 
As an important input into these calculations we make use of the Thomas-Fermi radius which we have measured in the second sound experiments (the measured Thomas-Fermi radii can be found in Supplementary Note 1). 
As an example, from the measurement at $(k_Fa)^{-1}=(1.91\pm0.05)$ we determine the peak superfluid fraction to be  $n_{s0}/n_0=0.98$ close to the trap center at maximum density, where the local $(k^{\mathrm{hom}}_Fa)^{-1}=(1.06\pm0.05)$ and  $T/T^{\mathrm{hom}}_C=(0.40\pm0.15)$, with $T^{\mathrm{hom}}_C=0.21T^{\mathrm{hom}}_F$ and $T^{\mathrm{hom}}_F=\hbar^2 (k^{\mathrm{hom}}_F)^2/2mk_B$.
For comparison, for a homogeneous weakly-interacting BEC with a superfluid fraction close to unity the temperature would need to be $T \ll T_C^{\mathrm{hom}}$, according to $n_s/n=1-\left(T/T^{\mathrm{hom}}_C\right)^{3/2}$. 
At unitarity, by contrast,  the superfluid fraction reaches unity already at $T/T^{\mathrm{hom}}_C \approx0.55$, as shown by Sidorenkov et al. \cite{Sid13}.  
As expected, this comparison shows that for a given $T/ T^{\mathrm{hom}}_C$ the superfluid fraction grows with interaction strength. 

\begin{figure}[hbp]
\includegraphics[trim=0 0 0 0,width=1\textwidth]{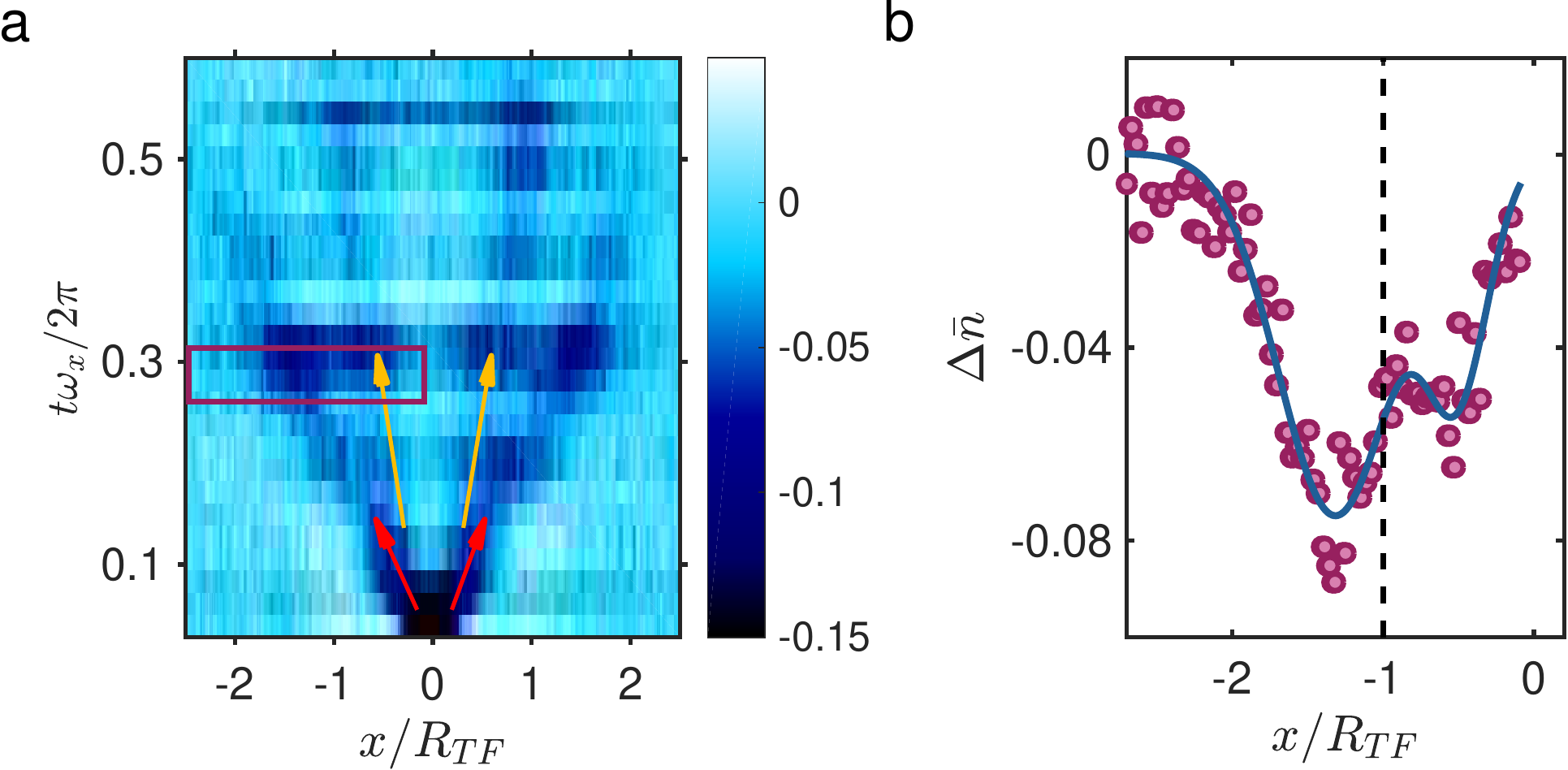}
\caption{\textbf{Comparing signal strength of first and second sound.}
\textbf{a,} Sound excitation experiment at $(k_Fa)^{-1}=(1.91\pm0.05)$ and at a temperature of $T/T_{\mathrm{C}}=(0.71\pm0.15)$. In contrast to Fig.\ref{fig:ComparisonTheoryExp}d, first sound (red arrows) and second sound (orange arrows) are now visible simultaneously. For $t\omega_x/2\pi<0.15$ first and second sound waves overlap and therefore cannot be distinguished from each other. 
\textbf{b,} shows $\Delta \bar n$ for $t=0.29\nu_x$. We fit the center position of each of the two sound waves using a Gaussian function (solid line).}
\label{fig:AmpDepSeries}
\end{figure}

\subsection*{Tuning the sound mode excitation}
In the following we investigate how the superfluid gas responds to different excitation protocols \cite{Hei06,Ara09,Hu10}. 
For this, we tune the excitation scheme, the excitation frequency and amplitude to gain additional insight into the nature of first and second sound.

\begin{figure}[hbp]
	\includegraphics[trim=150 50 150 0,width=.8\textwidth]{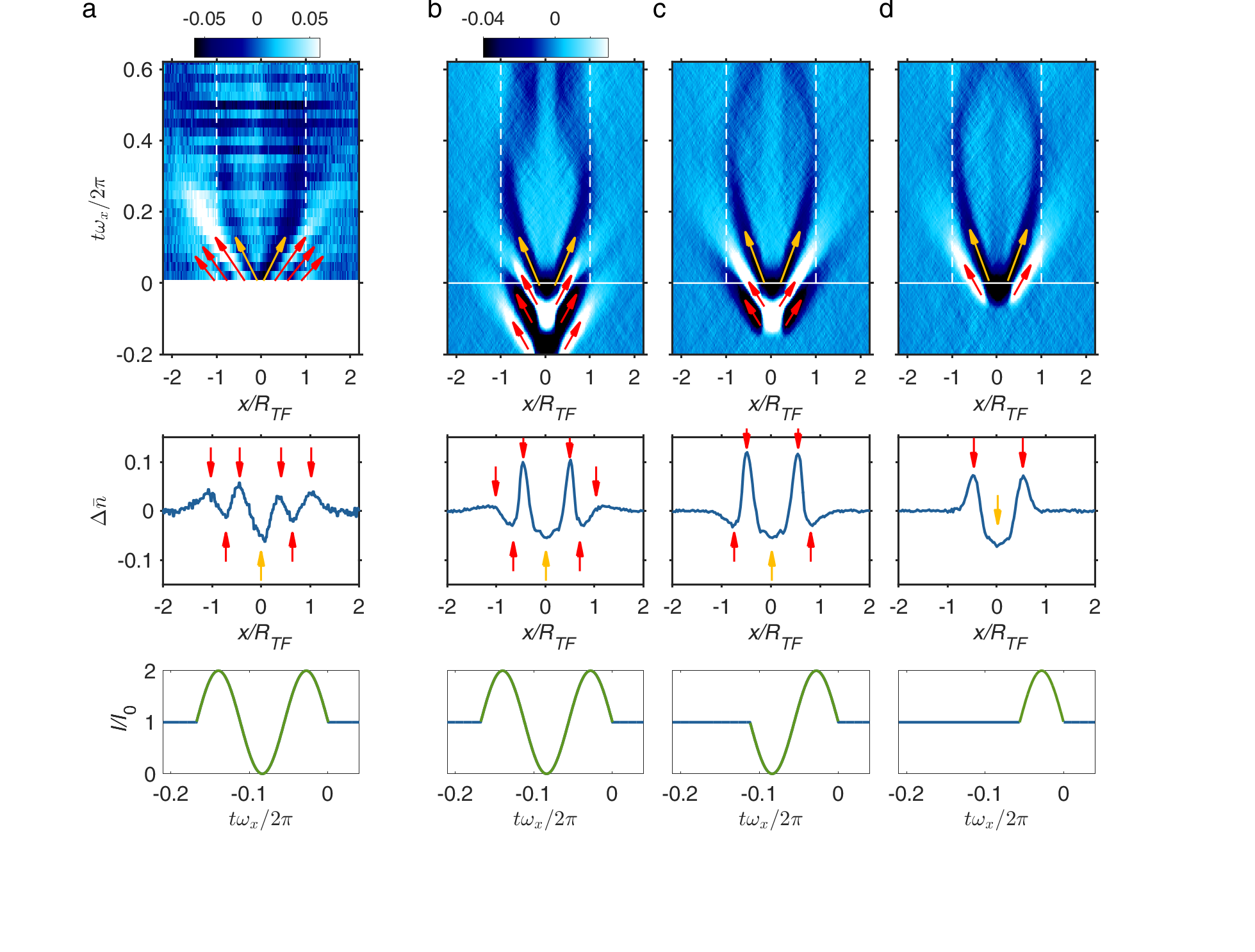}
	\caption{\textbf{Sound excitation with different modulation sequences.} \textbf{a,} $\Delta \bar n(x,t)$ data for $\omega_{ex}=0.61\omega_r$, $\Delta U =0.3\,U_0$ and at $\left(k_Fa\right)^{-1}=(1.91\pm0.05)$. The excitation pulse excites both, first and second sound waves (dark and bright traces). \textbf{b-d}, $\Delta \bar n(x,t)$ from numerical c-field simulations. Top row: False color images of $\Delta \bar n(x,t)$. First and second sound waves are marked with red and orange arrows, respectively. Mid row. Shown is $\Delta\bar{n}$ for $t=0$. Bottom row: Applied excitation scheme. } \label{fig:FreqDepSeries}
\end{figure}

In Fig.\,\ref{fig:AmpDepSeries}a we show the evolution of the system after a step pulse excitation at $B=735\,\mathrm{G}$ and $\Delta U=0.3\,U_0$, in which both, first and second sound are excited.
In contrast to the experiment in Fig.\,\ref{fig:ComparisonTheoryExp}d, the laser beam is abruptly switched off - not on.
As a consequence, the wave packets of both first and second sound now correspond to dips in the particle density.
In Fig. \ref{fig:AmpDepSeries}b we show the density distribution for the time and position range indicated by the purple rectangle in Fig. \ref{fig:AmpDepSeries}a. 
From a fit of two Gaussian dips to the two wave packets, we determine an amplitude ratio of $W_2/W_1\approx0.7$.
This result approximately matches the predictions of Ref.\cite{Ara09,Hu10} (see also Supplementary Note 3), where the response of both, a weakly and a strongly interacting molecular Bose gas has been derived.
The prediction yields $W_2/W_1=0.9$ for an interaction parameter of $(k_Fa)^{-1}=2$, which is of similar magnitude as our result. 

Next, we study the response for first and second sound waves after exciting them with short sinusoidal modulation sequences, as shown in Figs.\,\ref{fig:FreqDepSeries}a-d, where $\left(k_Fa\right)^{-1} = (1.91\pm0.05)$. 
The modulation frequency is $\omega_{ex}=0.61\omega_r$, so that parametric heating is somewhat suppressed and coupling to first sound is enhanced as compared to the experiment shown in Fig. \ref{fig:ComparisonTheoryExp}e.
The numerical simulations in Figs.\,\ref{fig:FreqDepSeries}b-d demonstrate how the excitation pattern produces a corresponding wave train of first sound. 
Once waves of first sound have propagated beyond the Thomas-Fermi radius they diffuse out and lose signal strength. 
The first sound wave train is always followed by a single dark second sound wave packet.
The experimental data in Fig.\,\ref{fig:FreqDepSeries}a agrees quite well with the simulation in Fig.\,\ref{fig:FreqDepSeries}b. 
Notably, the diffusion of the first sound wave train is somewhat less strong than in the simulations.
This descrepancy might be explained by the higher longitudinal trap frequencies used in the simulations which lead to a faster dispersion.

\section*{Conclusion}
In conclusion, we have studied second sound propagation in an ultracold Fermi gas of $^6$Li atoms across the BEC-BCS crossover for a range of different superfluidity at $T=0.7\,T_C$. 
We find the second sound velocity to vary only slightly across the BCS-BEC crossover, which is
in agreement with an interpolation of hydrodynamic theory \cite{Hei06}. 
In the BEC regime the results match numerical predictions based on c-field simulations.

Additionally, we investigate the response of the superfluid gas on various excitation pulse shapes, ranging from gentle local heating to an abrupt kick which allows for tuning waveform and amplitude of the sound modes. 
The responses of first and second sound are quite different, which hints at their different nature. 
We find that a particular useful excitation is a step wise excitation where both first and second sound propagate as  density dips. 
With this scheme we achieve similar amplitudes for second and first sound and the second sound wave can be easily distinguished from the first one.  
In the future it will be useful to extend our measurements in the strongly interacting regime to a larger range of temperatures below $T_C$. 
Since the second sound velocity is related to the local superfluid density, this measurement technique can help to construct the equation of state in the strongly interacting regime.
\newpage
\section*{Methods}

\subsection*{Calculating $\Delta \bar{n}$ from the density profiles}
Each of the experimental sound propagation images in Figs.\,1d-e, 3a, 4a is a time-ordered stack of one-dimensional column density profiles $\Delta\bar{n}(x,t)$ of the atom cloud.
A one-dimensional column density profile $n(x,t)$ is produced as follows:
For a given propagation time $t$ after the sound excitation ended we take an absorption image of a cloud to obtain the density distribution $n_{ex}(x,y,t)$. 
We integrate each absorption image along the y-axis to obtain a one-dimensional column density profile $n_{ex}(x,t)$. 
To reduce noise, we average 15 density profiles and obtain $\bar{n}_{ex}(x,t)$.
We repeat this procedure for an unperturbed cloud to obtain $\bar{n}(x)$. 
By subtracting the two density profiles from each other we obtain $\Delta \bar{n}(x,t)=(\bar{n}_{ex}(x,t)-\bar{n}(x))/\bar{n}(0)$.

\subsection*{Data availability}
The presented data are available from the corresponding author upon request.

\subsection*{Acknowledgements}
The authors thank Benjamin Deissler and Wladimir Schoch for the support in the stages of the experiment. 
Additionally, the authors thank Sandro Stringari, Hui Hu, Xia-Ji Liu, and Jia Wang for encouraging and illuminating discussions. 
V.P.S. and L.M. acknowledge the support from the DFG in the framework of SFB 925 and the excellence clusters 'The Hamburg Centre for Ultrafast Imaging'- EXC 1074 - project ID 194651731 and 'Advanced Imaging of Matter' - EXC 2056 - project ID 390715994. 
D.K.H., T.P., M.J., W.L. and J.H.D. acknowledge support from the Deutsche Forschungsgemeinschaft within SFB/TRR 21 (project part B4) and project LI988/6-1,  the  Baden-Württemberg  Foundation,  and  the  Center for Integrated Quantum Science and Technology (IQST).

\subsection*{Author contribution}D.K.H. and T.P. performed the experiments. 
D.K.H. and M.J. performed the data analysis, V.P.S. carried out analytic and numerical simulations. J.H.D and L.M. supervised the project. The manuscript was written by D.K.H., W.L., V.P.S., L.M. and J.H.D.

\subsection*{Competing interests}The authors declare no competing interests.

\subsection*{Corresponding author}Johannes Hecker Denschlag: johannes.denschlag@uni-ulm.de

\section*{Supplementary Information}
\renewcommand\thefigure{S\thesection\arabic{figure}}
\renewcommand\thetable{S\thesection\arabic{table}}

\newcolumntype{C}[1]{>{\centering\arraybackslash}p{#1}}
\setcounter{figure}{0}
\setcounter{equation}{0}

\subsection*{Supplementary Note 1: Temperatures to the measurements in Fig. 1 and Fig. 2}
In this section we present the temperatures to the measurements shown in Fig. 1 and Fig. 2 (see table \ref{tab:table-name}). 
We determine the temperatures by fitting a second order virial expansion of the density distribution at the wings of the cloud \cite{Pai18}. 
To compare the absolute temperature with $T_C$ for various interaction strengths we use values for $T_C$ as shown in figure \ref{fig:TTCcurve}. 

$T_C$ is not precisely known yet in the strongly interacting regime. 
In the limit of the BEC regime the BEC mean field model should give accurate values for critical temperature. 
Closer towards the resonance we expect the diagrammatic $t$-matrix calculation to provide quite good values \cite{Pin19}. 
For the range in between ($0.5 < (k_F a)^{-1} < 3$) we linearly interpolate between both $T_C$ curves. 

For the measurements on the BCS side we have compared our results with temperatures obtained from the approach in reference \cite{Luo09}, where the total energy and entropy of a cloud is measured for thermometry. 
We find reasonable agreement between the temperatures obtained from the two approaches.

\begin{table}[h]
\begin{tabular}{|c|c|c|c|c|c|}
\hline
$(k_F a)^{-1}$ & $T$ [nK] & $T/T_F$ & $T_C/T_F$ & $T/T_C$ & $R_{TF}$ [$\mathrm{\upmu}$m]\\
\hline
-0.26 $\pm$ 0.04 & 67 $\pm$ 23 & 0.12 $\pm$ 0.04 & 0.171 & 0.71 $\pm$ 0.24 & 110 $\pm$ 5\\
\hline
-0.09 $\pm$ 0.03 & 65 $\pm$ 22 & 0.12 $\pm$ 0.04 & 0.194 & 0.61 $\pm$ 0.21 & 115 $\pm$ 5\\
\hline
0 $\pm$ 0.02 & 114 $\pm$ 28 & 0.17 $\pm$ 0.05 & 0.207 & 0.81 $\pm$ 0.24 & 124 $\pm$ 5\\
\hline
0.16 $\pm$ 0.03 & 90 $\pm$ 30 & 0.17 $\pm$ 0.05 & 0.231 & 0.74 $\pm$ 0.22 & 139 $\pm$ 5\\
\hline
0.33 $\pm$ 0.04 & 90 $\pm$ 30 & 0.18 $\pm$ 0.05 & 0.256 & 0.69 $\pm$ 0.20 & 153 $\pm$ 5\\
\hline
0.45 $\pm$ 0.04 & 120 $\pm$ 30 & 0.22 $\pm$ 0.06 & 0.272 & 0.79 $\pm$ 0.22 & 156 $\pm$ 5\\
\hline
0.81 $\pm$ 0.05 & 120 $\pm$ 30 & 0.22 $\pm$ 0.06 & 0.316 & 0.69 $\pm$ 0.19 & 121 $\pm$ 5\\
\hline
1.22 $\pm$ 0.05 & 130 $\pm$ 30 & 0.24 $\pm$ 0.06 & 0.347 & 0.70 $\pm$ 0.17 & 108 $\pm$ 5\\
\hline
1.71 $\pm$ 0.05 & 150 $\pm$ 30 & 0.28 $\pm$ 0.06 & 0.379 & 0.73 $\pm$ 0.16 & 107$\pm$ 5\\
\hline
1.91 $\pm$ 0.05 & 140 $\pm$ 30 & 0.28 $\pm$ 0.06 & 0.391 & 0.71 $\pm$ 0.15 & 96 $\pm$ 5\\
\hline
\end{tabular}
\caption{\label{tab:table-name}\textbf{Temperatures and Thomas-Fermi radii to the measurements presented in Fig. 2 (main text).} The temperatures are given in nK as well as units of $T_F$ and $T_C$. For expressing the temperature in units of $T_C$ we use an interpolated critical temperature curve (see fig. \ref{fig:TTCcurve}).}
\end{table}

\subsection*{Supplementary Note 2: C-field simulation method}

Here we present our simulation method that is used to simulate sound mode dynamics in a condensate of $^{6}$Li molecules on the BEC side.
The system is described by the Hamiltonian
\begin{align} \label{eq_hamil}
\hat{H}_{0} = \int \mathrm{d}{\bf r} \Big[ \frac{\hbar^2}{2M}  \nabla \hat{\psi}^\dagger({\bf r}) \cdot \nabla \hat{\psi}({\bf r})  + V({\bf r}) \hat{\psi}^\dagger({\bf r})\hat{\psi}({\bf r}) + \frac{g}{2} \hat{\psi}^\dagger({\bf r})\hat{\psi}^\dagger({\bf r})\hat{\psi}({\bf r})\hat{\psi}({\bf r})\Big].
\end{align}
$\hat{\psi}$ and $\hat{\psi}^\dagger$ are the bosonic annihilation and creation operator, respectively. 
The 3D interaction parameter is given by $g=4\pi a_{dd} \hbar^2/M$, where $a_{dd}$ is the dimer-dimer scattering length and $M$ the dimer mass.
The external potential $V({\bf r})$ represents the cigar-shaped trap $V_{\mathrm{trap}}({\bf r})=M(\omega_{ax}^2x^2+\omega_r^2 r^2)/2$. 
$\omega_{ax}$ and $\omega_r$ are the axial and radial trap frequencies, respectively. 
$r=(y^2+z^2)^{1/2}$ is the radial coordinate.

To perform numerical simulations we discretize space with the lattice of $180 \times 35 \times 35$ sites and the discretization length $l= 0.5\, \upmu \mm$, where $l$ is chosen to be smaller than or comparable to the healing length $\xi$ and the de Broglie wavelength $\lambda$.
In our c-field representation we replace in Eq. \ref{eq_hamil} and in the equations of motion the operators $\hat{\psi}$ by complex numbers $\psi$, see Ref. \cite{Sin16}.
We sample the initial states in a grand-canonical ensemble of temperature $T$ and chemical potential $\mu$ via a classical Metropolis algorithm.
We obtain the time evolution of $\psi$ using the equations of motion.
We calculate the observables of interest and average over the thermal ensemble.
We use the trap frequencies $(\omega_{ax}, \omega_r ) = 2\pi \times (70 \, \mHz, 780 \, \mHz)$, which are higher than those in the experiments.
The reason for choosing higher $\omega_{ax}, \omega_r $ is that we need to keep the effective total lattice size small enough to be able to carry out the numerical calculations.
The scattering length $a_{dd}$ and the trap central density $n_0$ are the same as the experiments.
$a_{dd}$ varies in the range $a_{dd} = 720 - 1650 a_0$, where $a_0$ is the Bohr radius, and $n_0$ in the range $n_0 = 8.2 - 11.2 \, \upmu \mm^{-3}$.
This results in a cigar-shaped cloud of $N=4.0\times 10^{4} -4.5\times 10^{4}$ $^{6}$Li molecules.
The temperature varies in the range $T= 240 - 280\, \mnK$ or $T/T_c= 0.4 - 0.6$, where $k_\mB T_c \approx 0.94 \hbar (\omega_{ax} \omega_r^2 N)^{1/3}$ is the critical temperature of a noninteracting gas.

To excite sound modes we add the perturbation $\mathcal{H}_{ex}(t) = \int \mathrm{d}{\bf r}\, V({\bf r},t) n({\bf r})$, where $n({\bf r})$ is the density at the location  ${\bf r}=(x,y,z)$. The excitation potential $V({\bf r},t)$ is given by
\begin{equation}\label{eq_pot}
V({\bf r},t)  = V_0 (t) \exp \Bigl(- \frac{ (x-x_0)^2 + (z-z_0)^2}{2\sigma^2} \Bigr),
\end{equation}
where $V_0(t)$ is the time-dependent strength and $\sigma$ is the width. The location $x_0$, $z_0$ are chosen to be the trap center.
We excite sound modes following the scheme used in the experiment, where $\sigma$ and $V_0$ are chosen such that the changes in the local density due to the excitation potential are the same as in the experiment. 
We calculate the density profile $\bar{n}_{ex}(x, t)$, which is integrated in the radial direction. 
For sound propagation we examine $\Delta \bar{n}(x, t) = \bigl(\bar{n}_{ex}(x, t) - \bar{n}(x) \bigr)/\bar{n}(0) $, where $\bar{n}(x)$ is the density profile of the unperturbed cloud integrated in the radial direction and $\bar{n}(0)$ is the maximum density.

The time evolution of $\Delta \bar{n}(x, t)$ shows excitation of second sound identified by a vanishing sound velocity at $R_{TF}$.
We fit the density profile with a Gaussian to determine the second sound velocity $u_2$ at the trap center.
We note that $u_2$ changes only negligibly compared to the experimental errorbars for the temperatures in the range $T/T_c= 0.5 - 0.7$.

\subsection*{Supplementary Note 3: Analytic description of the sound modes}

In the following we present an analytic description of first and second sound based on the two-fluid hydrodynamic model for a uniform gas. 
The total density $n$ of the gas is a sum of the superfluid $n_s$ and normal fluid density $n_n$. 
The first and second sound mode squared velocities are  given by \cite{Pet08}

\begin{equation} \label{eq_sol}
u_{1/2}^2 = \frac{1}{2} (c_T^2 + c_2^2 + c_3^2) \pm \Bigl[ \frac{1}{4} (c_T^2 + c_2^2 + c_3^2)^2 - c_T^2 c_2^2  \Bigr]^{1/2},
\end{equation}

where $c_T^2 = 1/M(\partial p / \partial n)_T$ and $c_2^2 =   n_s s^2 T/(n_n c_V)$ representing the isothermal and entropic sound velocities, respectively.
$p$ is the pressure, $s$ the entropy per unit mass, $T$ the temperature, and $c_V = T ( \partial s/\partial T)_n$ the heat capacity per unit mass. 
The quantity $c_3^2  \equiv c_S^2 - c_T^2 = ( \partial s/\partial n)_T^2 (n^2 T/c_V)$ couples the sound velocities $c_2$ and $c_T$,
where $c_S^2 = 1/M(\partial p / \partial n)_s$ corresponds to the adiabatic sound velocity. 
The decoupled sound modes in the limit of vanishing $T$ are

\begin{align}\label{eq_sol_nocoup}
u_1^2 = c_T^2 = \frac{1}{M}  \Bigl(  \frac{ \partial p}{\partial n}  \Bigr)_T \quad \text{and} \quad
u_2^2 = c_2^2 =  \frac{ n_s }{n_n} \frac{ s^2 T}{c_V} .
\end{align}
Here, first and second sound can be described as a pressure and entropy wave, respectively.
To determine the second sound velocity $u_2$, we calculate the entropy and the normal fluid density defined as

\begin{align}\label{eq_ent}
S = \sum_{\bk} \Big( -f_k \log f_k \pm (1 \pm f_k ) \log (1 \pm f_k )  \Bigr)
\end{align}

and 

\begin{align} \label{eq_rhon}
n_n =\frac{1}{M} \int \frac{d\mathbf{k}^3} {(2\pi)^3 } \, \frac{\hbar^2 k^2}{3} \Bigl( -  \frac{\partial f_k}{ \partial E_k} \Bigr),
\end{align}

respectively \cite{Pet08}.
$f_k = 1/\bigl( \exp(E_k/k_\mB T) \mp 1 \bigr)$ is the thermal occupation number, where $E_k$ is the excitation energy and $\bk$ the wavevector. 
The upper and lower sign correspond to a Bose and Fermi gas, respectively.

\subsection{BEC}

We use the Bogoliubov theory, valid in the dilute limit, to analyze the regime $k_\mB T<gn$, where $gn$ is the mean-field energy. 
The Bogoliubov spectrum is given by $E_k = \sqrt{\epsilon_k (\epsilon_k + 2 gn)}$, where $\epsilon_k= \hbar^2 k^2/(2M)$ is the free-particle spectrum. 
$M$ is the molecular mass.
To examine the decoupled modes in Eq. \ref{eq_sol_nocoup} we approximate $E_k$ by the linear spectrum $E_k \approx \hbar c k$, where $c= \sqrt{gn/M}$ is the Bogoliubov sound velocity. 
We obtain the entropy and the normal fluid density, respectively, 
\begin{align}
S = V \frac{2 \pi^2 }{45 \hbar^3} (k_\mB T)^3 \Bigl( \frac{M }{ gn} \Bigr)^{3/2} \quad \text{and} \quad 
n_n = \frac{2 \pi^2}{45} \frac{ (k_\mB T)^4}{\hbar^3}  \frac{M^{3/2}}{\left(gn\right)^{5/2}}.
\end{align}
The entropy per unit mass is $s = S/(NM) = g n_n/(M T)$ and the heat capacity per unit mass is $c_V = 3 s$.

Within upper description we can deduce following sound speeds

\begin{align}\label{eq_bec_u2}
u_1 = \sqrt{ \frac{ gn }{ M }}  \quad \text{and} \quad  
u_2 = \sqrt{\frac{1}{3}  \frac{ gn}{M }}.
\end{align}

Here, $u_2$ is $u_1/\sqrt{3}$. 
This result is only valid at zero temperature, see Fig. \ref{Fig1}a, where we show the full numerical solutions of Eq. \ref{eq_sol} using the Bogoliubov description. 

For $k_\mB T> gn$ instead we make use of a thermal gas description to determine $s$, $c_V$, and $n_n$, which are given by $s=2.568 k_\mB n_n/(2M n)$, $c_V=3s/2$, and $n_n= n (T/T_{\mathrm{C}})^{3/2}$, respectively\cite{Pet08}. 
In our experiments on the BEC side $k_\mB T/gn$ ranges from 1.9 to 3.2 which allows us to apply the thermal gas description.  

In this regime, solving eq. \ref{eq_sol} the sound velocities read,

\begin{align}
u_1 = \sqrt{ \frac{gn}{M} + \frac{ 0.856 k_\mB T}{M} }  \quad  \text{and}  \quad 
u_2 = \sqrt{ \frac{n_s}{n} \frac{gn}{M} }.
\label{eq:FirstSecondBEC}
\end{align}

$u_2$ is proportional to $\sqrt{n_s/n}$ and can be approximated by $u_2 = \sqrt{\bigl(1 - (T/T_{\mathrm{C}})^{3/2} \bigr) gn/M }$ (see Fig. \ref{Fig1}a).

\subsection*{Sound amplitudes}
Besides the sound velocity, our analytic description can be used to determine the amplitudes of the propagating sound modes, described as \cite{Ara09}

\begin{equation}
\delta n(x,t) = W_1 \delta\tilde{n}(x\pm u_1 t) + W_2 \delta \tilde{n}(x \pm u_2t).
\end{equation}
 
where $\delta\tilde{n}(x, t)$ is the density variation created by the excitation potential. 
$\delta\tilde{n} (x \pm u_{1/2} t)$ represent wave packets of first and second sound with weights $W_{1/2}$. The relative weight is given by
 
%
\begin{align} \label{eq_w2w1}
\frac{W_2}{W_1} = \frac{c_2^2 - u_2^2}{u_1^2 - c_2^2 } \frac{u_1^2}{u_2^2}
\end{align}

We determine $W_2/W_1$ by numerically solving Eq. \ref{eq_sol} for the regimes $k_\mB T < gn$ and $k_\mB T > gn$ using the Bogoliubov and thermal gas description, respectively. 

\begin{figure}[htp]
\centering
\includegraphics[width=0.85\textwidth]{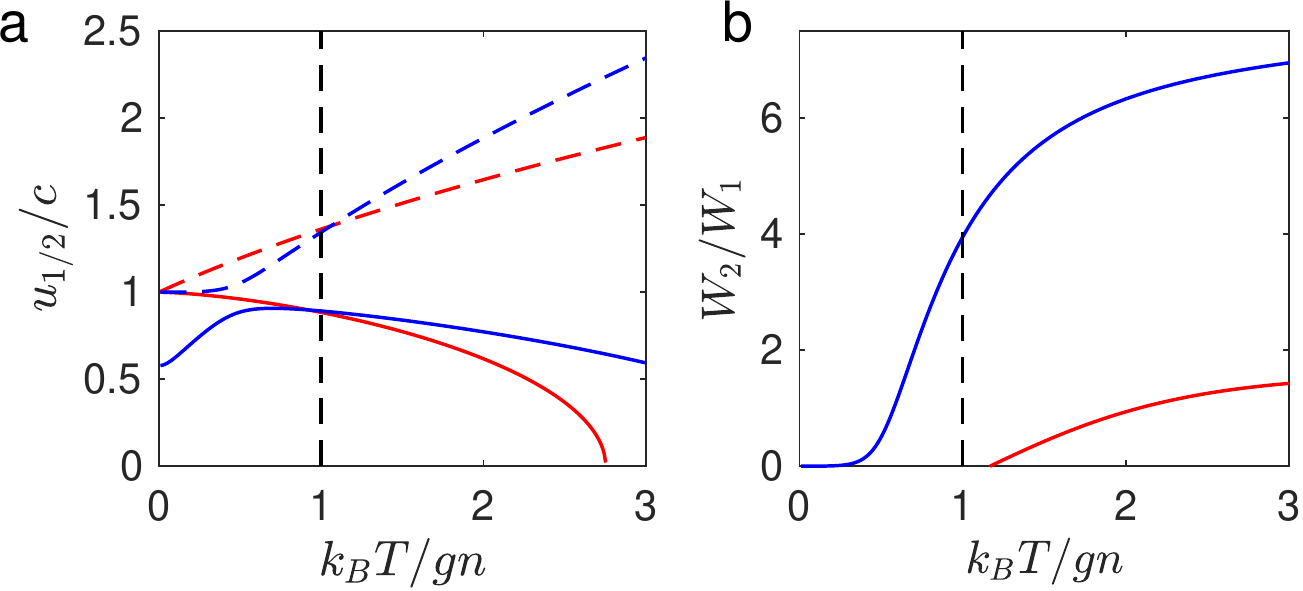}
\caption{\textbf{Sound velocities and amplitudes. a,} Sound velocities $u_{1/2}$  are determined from eq. \ref{eq_sol} and are shown as a function of $k_\mB T/gn$ using the Bogoliubov (blue lines) and thermal gas description (red lines). Here, $c$ is the Bogoliubov sound speed introduced in the text.
\textbf{b,} shows the relative weight $W_2/W_1$ for $k_\mB  T< gn$ (blue line) and $k_\mB T > gn$ (red line). 
}
\label{Fig1}
\end{figure}

\begin{figure}[htp]
	\centering
  \includegraphics[width=0.7\textwidth]{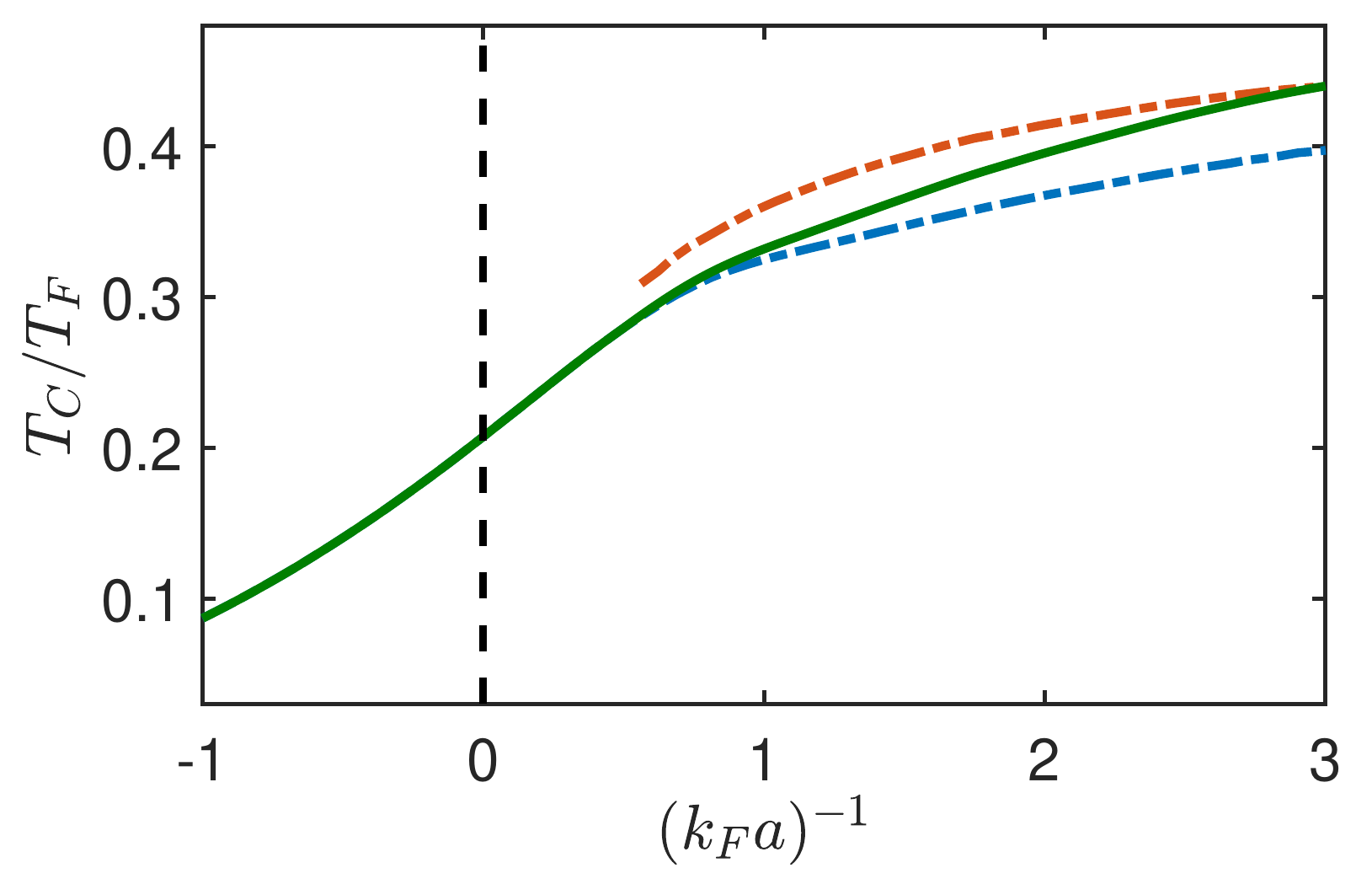}
  \vspace{-5mm}
\caption{\textbf{Critical temperature $T_C$ in units of $T_F$ as a function of $(k_Fa)^{-1}$ for a harmonically trapped Fermi gas.} The blue dash-dotted line shows a diagrammatic $t$-matrix calculation and the orange dash-dotted line a calculation based on a BEC mean field model  \cite{Pin19}. The green straight line interpolates linearly between the two approaches.}
	\label{fig:TTCcurve}
\end{figure}

We show these results in Fig. \ref{Fig1}b. The Bogoliubov description of the weight works only for $k_BT\ll gn$. 
We note that at higher temperatures terms beyond Bogoliubov are needed to account for the thermal damping of the modes. 
The Bogoliubov description thus leads to an overestimation of the weight at high temperatures.   
For temperatures above the mean-field energy the weight is described by the thermal gas description, which we use to estimate the relative weight of the two modes in the main text. 
Please note that the thermal description gives unphysical solutions for $k_BT/gn\rightarrow1$. 

\subsection{BCS} \label{sec_bcs}

A condensate of an interacting Fermi gas is described by the BCS spectrum $E_k = \sqrt{ \xi_k^2 + \Delta^2 }$, with $\xi_k = \hbar^2 k^2/(2m) - \mu$, where $\mu$ is the chemical potential and $\Delta (T)$ the gap.
At low $k_\mB T \ll \Delta $, we use $\mu \approx E_F$ and expand $\xi_k$ near the Fermi surface, i.e. $\xi_k = \hbar^2 k^2/(2m) - E_F \approx \hbar v_F |k -k_F|$ (see Ref. \cite{Lif80}). 
The entropy in Eq. \ref{eq_ent} results in
\begin{align}\label{eq_s_bcs}
S = \frac{3 N_\mtot}{E_F} \int_0^\infty d \xi_k \, \frac{E_k}{k_\mB T} \exp \Bigl(- \frac{E_k}{k_\mB T} \Bigr) = 3 N_\mtot \frac{\Delta_0}{E_F} \sqrt{ \frac{\pi \Delta_0}{ 2 k_\mB T} }  \exp \Bigl(- \frac{\Delta_0 }{k_\mB T} \Bigr), 
\end{align}
with
\begin{align} \label{eq_gap0}
\Delta_0 = (2/e)^{7/3} E_F \exp\big(\pi/(2k_F a) \bigr)
\end{align}
which is the gap at zero temperature \cite{Gor61}. With Eq. \ref{eq_s_bcs} we determine $s = S/(m N_\mtot)$ and $c_V$. 
The normal fluid density in Eq. \ref{eq_rhon} gives

\begin{align}
\frac{n_n}{n_\mtot} = 2  \int_0^\infty d \xi_k \,  \Bigl( - \frac{ \partial f_k }{ \partial E_k  } \Bigr) =  \sqrt{ \frac{2\pi \Delta_0}{ k_\mB T} } \exp \Bigl(- \frac{\Delta_0 }{k_\mB T} \Bigr).
\end{align}
 
Using $s$, $c_V$, and $n_n $ in Eq. \ref{eq_sol_nocoup} we obtain the second sound velocity

\begin{align}\label{eq_bcs_u2}
u_2 = \frac{ \sqrt{3}}{2} \frac{k_\mB T }{E_F} v_F,
\end{align}

which is valid for $T<T_C$.
The BCS critical temperature is given by $k_\mB T_C = (\gamma/\pi) \Delta_0 = 0.567 \Delta_0$, which depends on the interaction parameter $(k_F a)^{-1}$.
We show in the main text the result $u_2$ at various interactions on the BCS side (see Fig. 2).
$u_2$ vanishes at zero temperature contrary to the BEC superfluids. 
We note that this result is consistent with Ref. \cite{Hei06}.

\subsection*{Supplementary Note 4: BEC mean-field model}
To estimate the density distribution of a partially Bose condensed cloud in the BEC regime we carry out a self-consistent calculation where the condensate phase is treated  within the Thomas-Fermi approximation and for the normal phase we use a standard thermodynamical approach. 
Specifically, we solve the following set of coupled equations \cite{Pit03} 

\begin{equation}
n_\text{s}(\textbf{r}) = \frac{\mu_\text{s} - V_\text{ext}(\textbf{r}) - 2 g n_\text{n}(\textbf{r}) }{g} \Theta\left( \mu_\text{s} - V_\text{ext}(\textbf{r}) - 2 g n_\text{n}(\textbf{r})\right)\label{eq:TFeq}
\end{equation}
\begin{equation}
n_\text{n}(\textbf{r}) = \frac{1}{\lambda_{dB}^3} \text{Li}_{3/2} \left(  \exp \left\lbrace \frac{\mu_\text{n} - V_\text{ext}(\textbf{r}) - 2 g n_\text{s}(\textbf{r}) - 2 g n_\text{n}(\textbf{r})}{k_B T} \right\rbrace \right).\label{eq:PartDist}
\end{equation}

Here, $\lambda_\text{dB}$ is the thermal deBroglie wavelength, $g = 4 \pi \hbar^2 a_{dd} /M$ is the coupling constant, $T$ is the temperature and $V_\text{ext}(\textbf{r})$ is the external potential consisting of the harmonic trapping potential and the repulsive potential of the excitation beam, $\mu_\text{s}$ and $\mu_\text{n}$ are the chemical potentials of the superfluid and the normal fluid part, respectively. 
For the calculation we set $\mu_\text{n} = \min[V_\text{ext}(\textbf{r}) + 2 g n_\text{s}(\textbf{r}) + 2 g n_\text{n}(\textbf{r})]$ which ensures that the normal gas reaches the critical density $n_\text{n,crit} = \text{Li}_{3/2}(1)/\lambda_\text{dB}^3$ at the Thomas-Fermi radius. 
This way, the number of normal fluid atoms is fixed. 
$\mu_\text{s}$ is chosen such that the total atom number matches the experimental value. 

\begin{figure}[htp]
	\centering
  \includegraphics[width=0.9\textwidth]{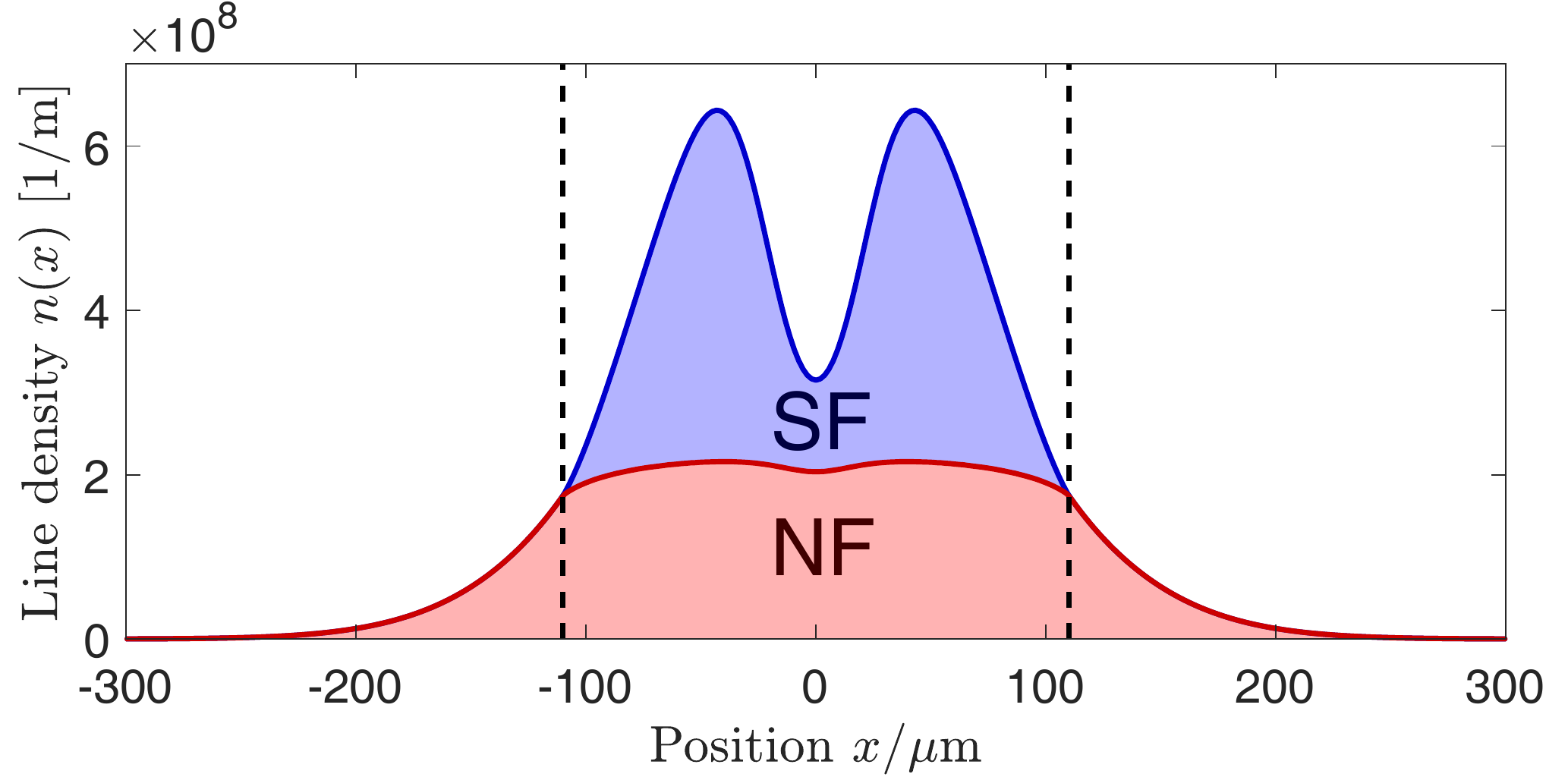}
\caption{\textbf{Axial line densities of the superfluid and the normal phase obtained from a self-consistent calculation}. The calculation is performed for a gas at $(k_Fa)^{-1}=1.91$ and a temperature of $T=145\,\mathrm{nK}$. The repulsive potential of the excitation laser beam at the center locally reduces the density of the cloud. The vertical dotted lines indicate the Thomas-Fermi radius at $x = \pm 110\, \mathrm{\upmu}$m.}
	\label{fig:Densities}
\end{figure}

Equation \ref{eq:TFeq} represents the Thomas-Fermi approximation where we take into account the repulsive mean-field potential of the normal fluid part. Equation \ref{eq:PartDist} is the density distribution of a thermal bosonic cloud, again including the additional mean field potential produced by the atoms. 
By self-consistently solving the coupled equations we obtain the density distributions of the superfluid and the normal fluid gas as shown in fig. \ref{fig:Densities}.

\end{document}